\documentclass[aps,amsmath,amssymb,superscriptaddress,pra]{revtex4-1}   
\usepackage[latin2]{inputenc}
\usepackage{amsmath,amssymb}
\usepackage{amsfonts}
\usepackage{bm}
\usepackage{dcolumn}
\usepackage{setspace}
\usepackage{graphicx}
\usepackage{color}
\usepackage{multirow}
\usepackage{verbatim}
\usepackage{epstopdf}

\def\bar{\begin{array}}
\def\ear{\end{array}}

\def\f{\frac}
\def\M{\mathcal{I}}
\def\R{\mathbf{R}}
\def\r{\mathbf{r}}

\def\nn{\nonumber}

\def\M{\mathcal{M}}

\def\p{\partial}

\def\Mm{\mathrm{Im}}
\def\o{\overline}

\def\oQ{\overline{Q}}

\begin{document}

\title{Gauge- and coordinate-invariant equations for two-component systems}

\author{Ryan Requist} 
\affiliation{Fritz Haber Center for Molecular Dynamics, Institute of Chemistry, The Hebrew University of Jerusalem, Jerusalem 91904 Israel}

\date{\today}

\begin{abstract}

The Schr\"odinger-like equations for the marginal and conditional probability amplitudes resulting from the exact factorization of the wavefunction of a two-component system are derived in a form that is invariant to gauge and coordinate transformations.  Coupled equations equivalent to the nonrelativistic Schr\"odinger equation of a molecule are derived as an example.

\end{abstract}

\maketitle

\section{Introduction}

In calculations of systems containing two types of particles, one can benefit by exploiting any differences between the particles.  The quintessential example is the Born-Oppenheimer (BO) approximation, which takes advantage of the large disparity in the masses of electrons and nuclei to adiabatically decouple the effective equations describing their motion.
It starts from the Ansatz $\Psi(r,R)=\chi(R) \Phi_R(r)$ for the full wavefunction, written as the product of a marginal nuclear wavefunction $\chi(R)$ and a conditional electronic wavefunction $\Phi_R(r)$ depending parametrically on the $N$-tuple of nuclear coordinates $R=\{\R_1,\cdots,\R_N\}$.

The factorization introduces gauge freedom due to the possibility of simultaneously redefining $\chi(R)\rightarrow e^{i\lambda(R)}\chi(R)$ and $\Phi_R(r)\rightarrow e^{-i\lambda(R)}\Phi_R(r)$.  
Maintaining gauge invariance in the effective Schr\"odinger equation for the nuclei,
\begin{align}
\sum_{n=1}^N \f{(\mathbf{P}_{n}+\mathbf{A}_{n})^2}{2M_{n}} \chi(R) + \mathcal{E}(R) \chi(R) = E \chi(R) 
{,} \label{eq:BO:lab:chi}
\end{align}
requires that it contain 
a vector potential $\mathbf{A}_{n}=-i\langle \Phi_R | \nabla_{n} \Phi_R\rangle$ ($\nabla_n\equiv\nabla_{\mathbf{R}_{n}}$), the Mead-Truhlar vector potential \cite{mead1979}, 
which obeys the same gauge transformation law as an electromagnetic vector potential.  The notable difference is that the gauge freedom is larger, since the phase $\lambda(R)$ is an arbitrary function of $N$ nuclear coordinates rather than merely a function on three-dimensional space.

Recently, an approach that goes beyond the adiabatic BO decoupling 
has been developed \cite{gidopoulos2014}.
Since it relies on the idea that the exact $\Psi(r,R)$ can be factorized into a marginal nuclear probability amplitude and a conditional electronic probability amplitude \cite{hunter1975}, it is known as the {\it exact factorization (EF) method}.  
The Schr\"odinger equation for the marginal wavefunction $\chi(R)$ has the same form as Eq.~(\ref{eq:BO:lab:chi}), but 
the Schr\"odinger-like equation for the electronic wavefunction $\Phi_R(r)$ is no longer an eigenvalue equation, as it was in the BO approximation.

A new feature appears in the time-dependent extension of the EF method \cite{abedi2010}: the potential energy surface $\mathcal{E}(R,t)$ and vector potential $\mathbf{A}_n(R,t)$ become time dependent, in contrast to the time-dependent BO approximation, where they remain static quantitites.  Thus, the time-dependent EF method has a larger degree of gauge freedom as one is free to perform a gauge transformation with a time-dependent phase, $\lambda(R,t)$. 
The gauge potentials continue to transform analogously to the electromagnetic gauge potentials under space- and time-dependent gauge transformations.

Approximations are needed to 
make the EF method into a practical computational scheme.  
Mixed quantum-classical methods based on the EF equations \cite{agostini2013,abedi2014,min2015,agostini2016}, and using an ensemble of classical trajectories to sample the nuclear wavepacket, have been applied to 
excited state molecular dynamics and decoherence
\cite{agostini2015,min2017,curchod2018,villaseco-arribas2022}. 
For large molecules and solids, EF-based density functional theory \cite{requist2016b,li2018} justifies bypassing the correlated electronic EF equation in favor of
Kohn-Sham-like equations.
A local conditional density approximation correctly describes electron transfer in the LiF molecule \cite{li2018}.  
While electron-phonon interaction effects are usually built on top of a BO ground state calculation through density functional perturbation theory \cite{gonze1997a,gonze1997b,baroni2001}, 
EF-based density functional theory encompasses electronic and phononic degrees of freedom in an overarching variational framework \cite{requist2019b}.
Nonadiabatic effects such as the phonon-induced wiggle in photoemission spectroscopy are captured by an orbital-dependent nonadiabatic functional approximation \cite{requist2019b,pellegrini2022}.
Geometric phase effects can be incorporated in BO molecular dynamics through a semiclassical closure of the Ehrenfest equations \cite{rawlinson2020}.  The resulting regularization of conical intersection-induced singularities in the gauge potential reproduces features observed in nonadiabatic calculations \cite{requist2016a}.

The EF concept is not limited to two-component systems of electrons and nuclei.  The marginal and conditional amplitudes can in fact describe the same species of particle.
For example, an $n$-body electronic wavefunction can be factorized into a one-body marginal wavefunction and an $(n-1)$-body conditional wavefunction \cite{hunter1986,schild2017}.  Atomic strong-field dynamics and ionization \cite{schild2017,kocak2020} as well as the charge transfer steps in 
density functional theory \cite{kocak2021} have been studied from this point of view.
In a system of electrons, nuclei and photons described within the dipole approximation, the EF method has been used to define an exact Schr\"odinger-like equation for the photons and characterize
the deviations of its time-dependent effective potential from the bare quadratic potential of a matter-free system \cite{hoffmann2018}.
The extension of the exact factorization concept to states living in the direct product of any two Fock spaces \cite{gonze2018}
has been invoked to propose embedding theories for strongly interacting electronic systems \cite{lacombe2020,requist2021}.

Due to limitations on computational resources, it is often necessary to restrict nuclear configuration space to a reduced manifold of collective nuclear coordinates.  Generally, such collective coordinates are taken to be curvilinear.  Since the original laboratory frame EF equations, containing the Laplacian in Cartesian coordinates, are not invariant under general coordinate transformations, the nuclear kinetic energy operator has to be rederived for each new set of coordinates.
In this paper, I derive the EF equations of a two-component system in a form that is invariant to general gauge and coordinate transformations, underscoring the role of Riemannian geometry in the EF equations.  
In Sec.~II, the gauge invariance of the EF equations is reviewed.  The coordinate-invariant EF equations are derived in Sec.~III, and the Riemannian objects they contain are identified.  The EF equations also contain a new geometric object, called the quantum Christoffel symbol, which is discussed in Sec.~IV.  Conclusions are given in Sec.~V.

\section{Gauge invariance of the exact factorization equations}

In this section, we review the gauge invariance of the time-dependent EF equations \cite{abedi2010} from the point of view of the covariant derivative.
The time-independent EF equations are also known to be gauge invariant \cite{gidopoulos2014}.

To be concrete, we consider a system of electrons and nuclei with laboratory frame coordinates $r=(\r_1,\r_2,\ldots)$ and $R=(\R_1,\ldots,\R_N)$ as representative of a general two-component system.  The Hamiltonian of the system is
\begin{align}
\hat{H} = \sum_n \f{\mathbf{P}_n^2}{2M_n} + \sum_i \f{\mathbf{p}_i^2}{2m} + \hat{V}_{ee} + \hat{V}_{nn} + \hat{V}_{en}  {,}
\end{align}
where $\hat{V}_{ee}$, $\hat{V}_{nn}$, and $\hat{V}_{en}$ are the electron-electron, and nucleus-nucleus, and electron-nucleus Coulomb repulsions, respectively.
Adapting the notations of Ref.~\cite{gidopoulos2014}, we write the laboratory frame time-dependent EF equations as
\begin{align}
(i\hbar \p_t - A_0) \chi(R,t) &= \sum_n \f{(\mathbf{P}_n+\mathbf{A}_n)^2}{2M_n} \chi(R,t) + \mathcal{E}(R,t) \chi(R,t) \label{eq:EF:lab:chi} \\
(i \hbar \p_t +A_0) \Phi_R(r,t) &= \sum_n \f{(\mathbf{P}_n-\mathbf{A}_n)^2}{2M_n} \Phi_R(r,t) + \sum_n \f{1}{M_n} \f{(\mathbf{P}_n+\mathbf{A}_n)\chi}{\chi} \cdot (\mathbf{P}_n-\mathbf{A}_n) \Phi_R(r,t) \nn \\
&\quad+ \big[\hat{H}^{BO} - \mathcal{E}(R,t)\big] \Phi_R(r,t) {,}
\label{eq:EF:lab:Phi}
\end{align}
where $A_0 = -i \hbar \langle \Phi _R| \p_t \Phi_R\rangle$, $\mathbf{A}_{n}=-i\hbar\langle \Phi_R | \nabla_{n} \Phi_R\rangle$, $\mathbf{P}_{n}=-i\hbar\nabla_{n}$, $\hat{H}^{BO}=\hat{H}-\sum_n \mathbf{P}_n^2/2M_n$, and
\begin{align}
\mathcal{E}(R,t) = \langle \Phi_R | \hat{H}^{BO} | \Phi_R\rangle + \sum_n \f{\langle (\mathbf{P}_n-\mathbf{A}_n) \Phi_R | (\mathbf{P}_n-\mathbf{A}_n) \Phi_R \rangle}{2M_n} {.}
\end{align}
To see that the equations are invariant under the gauge transformation
\begin{align}
\chi(R,t) &\rightarrow e^{-i\lambda(R,t)} \chi(R,t) \nn \\
\Phi_R(r,t) &\rightarrow e^{i\lambda(R,t)} \Phi_R(r,t) {,}
\label{eq:gauge}
\end{align}
we first observe that the gauge potentials transform as 
\begin{align}
A_0 &\rightarrow A_0 + \hbar \p_t \lambda \nn \\
\mathbf{A}_{n} &\rightarrow \mathbf{A}_{n} + \hbar \nabla_{n} \lambda {.}
\label{eq:gauge:potentials}
\end{align}
It follows that $(\mathbf{P}_n+\mathbf{A}_n)\chi$ and $(\mathbf{P}_n-\mathbf{A}_n)\Phi_R$ transform in the same way as $\chi$ and $\Phi_R$, respectively, i.e.
\begin{align}
(\mathbf{P}_n+\mathbf{A}_n)\chi(R,t) &\rightarrow e^{-i\lambda(R,t)} (\mathbf{P}_n+ \mathbf{A}_n)\chi(R,t) \nn \\
(\mathbf{P}_n-\mathbf{A}_n)\Phi_R(r,t) &\rightarrow e^{i\lambda(R,t)} (\mathbf{P}_n- \mathbf{A}_n)\Phi_R(r,t) {,}
\end{align}
and similarly for $(i\hbar \p_t - A_0) \chi$ and $(i\hbar \p_t + A_0)\Phi_R$.  Since quadratic terms like $(\mathbf{P}_n+\mathbf{A}_n)^2\chi$ also transform in the same way, it is evident that under a gauge transformation every term in Eq.~(\ref{eq:EF:lab:chi}) is merely multiplied by $e^{-i\lambda}$, while every term in Eq.~(\ref{eq:EF:lab:Phi}) is multiplied by $e^{i\lambda}$. 

It is precisely the presence of the gauge potentials that allows us to differentiate objects that are not uniquely defined on account of their gauge freedom.
The gauge-covariant derivatives 
\begin{align}
\mathbf{D}_{n} &= \nabla_{n} + \f{i}{\hbar} \mathbf{A}_n \nn \\
D_t &= \p_t + \f{i}{\hbar} A_0 \label{eq:gauge-covariant:derivative}
\end{align}
can be used to differentiate any function that obeys the same transformation law as $\chi$.
The gauge-covariant derivatives needed for differentiating functions that transform like $\Phi_R$ are $\mathbf{D}_{n} = \nabla_{n} - \f{i}{\hbar} \mathbf{A}_n$ and $D_t = \p_t - \f{i}{\hbar} A_0$.
Although we use the same symbols 
for different derivatives, 
which one is meant will always be clear from the context, namely according to which type of function the derivative acts on.
The derivatives in Eq.~(\ref{eq:gauge-covariant:derivative}) are 
It is now evident that every derivative in Eqs.~(\ref{eq:EF:lab:chi}) and (\ref{eq:EF:lab:Phi}) is a gauge-covariant derivative.  Indeed, a necessary condition for the EF equations to be meaningful is that it is possible to write them solely in terms of gauge-covariant derivatives.

The above-described gauge geometry of the nuclear BO Schr\"odinger equation in Eq.~(\ref{eq:BO:lab:chi}) has proven to be important because it is directly connected with molecular geometric phase effects.
Just as the electromagnetic vector potential, even if its curl is zero throughout the region where a charged particle moves, gives rise to observable effects through the Aharonov-Bohm effect \cite{aharonov1959}, 
the Mead-Truhlar vector potential $\mathbf{A}_{n}^{BO} = -i \langle \Phi_R^{BO} | \nabla_n \Phi_R^{BO}\rangle$ generates 
a molecular Aharonov-Bohm effect \cite{mead1980a} via the phase factor
\begin{align}
e^{\f{i}{\hbar} \int \sum_n \mathbf{A}_{n}^{BO}\cdot d\mathbf{R}_{n}} {.}
\end{align}
In molecules with real-valued Hamiltonians, this phase factor is an alternative way of viewing the topological sign change of the electronic wavefunction 
discovered by Longuet-Higgins \cite{longuet-higgins1958}.  For complex-valued Hamiltonians, e.g.~when spin-orbit interactions are turned on \cite{mead1980b}, the phase factor is a $U(1)$ holonomy, a molecular realization of the quantum mechanical geometric phase \cite{berry1984}.

The nuclear EF equation in Eq.~(\ref{eq:EF:lab:chi}) has exactly the same form as the BO equation in Eq.~(\ref{eq:BO:lab:chi}), and as seen in the gauge transformation rules in Eq.~(\ref{eq:gauge:potentials}), 
it exhibits gauge geometry like the BO equation.  However, the EF equations in Eqs.~(\ref{eq:EF:lab:chi}) and (\ref{eq:EF:lab:Phi}), and the gauge potentials they contain, do not depend on any adiabatic approximation.  Hence, the molecular geometric phase defined in terms of them is an exact quantity \cite{gidopoulos2014,abedi2010}.  There are cases where the adiabatic geometric phase is $\pi$ but the exact geometric phase is zero \cite{min2014}.  In such cases, the adiabatic geometric phase is an artifact of the BO approximation.  On the other hand, in pseudorotating triatomic molecules, Jahn-Teller defects and other systems with degeneracies, 
the exact geometric phase is nonzero for current-carrying eigenstates \cite{requist2016a}.

\section{Coordinate-invariant exact factorization equations}

By performing the linear coordinate transformation $(r,R)\rightarrow (q,Q,\R_{cm})$, where $\R_{cm}$ is the center of mass and the electronic coordinates $q=(\mathbf{q}_1,\mathbf{q}_2,\ldots)$ and the $N-1$ generalized nuclear coordinates $Q^{\mu}$ are translation-invariant, the time-independent EF equations were derived in the form \cite{requist2016b}
\begin{align}
&\f{1}{2} \M^{\mu\nu} (P_{\mu}+A_{\mu})(P_{\nu}+A_{\nu}) \chi + (\mathcal{E}^{BO} + \mathcal{E}_{geo}) \chi = E \chi \label{eq:EF:cm:chi} \\
&\f{1}{2} \M^{\mu\nu} (P_{\mu}-A_{\mu})(P_{\nu}-A_{\nu}) |\Phi\rangle + \M^{\mu\nu} \f{(P_{\mu}+A_{\mu})\chi}{\chi} (P_{\nu}-A_{\nu}) |\Phi\rangle + \hat{H}^{BO} |\Phi\rangle = (\mathcal{E}^{BO} + \mathcal{E}_{geo}) |\Phi\rangle{,}
\label{eq:EF:cm:Phi}
\end{align}
where $P_{\mu}=-i\hbar\partial/\partial Q^{\mu}$, $A_{\mu}=\langle \Phi | P_{\mu} \Phi\rangle$, $\mathcal{E}^{BO} = \langle \Phi | \hat{H}^{BO} | \Phi \rangle$, $\mathcal{E}_{geo} = \f{1}{2}\M^{\mu\nu} \mathrm{Re} \langle (P_{\mu}-A_{\mu}) \Phi| (P_{\nu}-A_{\nu}) \Phi\rangle$ and $\M^{\mu\nu}$ is a $Q$-independent tensor that is the inverse of the mass (inertia) tensor $\M_{\mu\nu}$.  Here and in the following we use a mixed representation where the marginal amplitude of the nuclei is represented as a wavefunction $\chi(Q)$ on configuration space and bra-kets are used to represent the electronic state.  The motivation for using this mixed representation is that the electronic state is usually described in terms of a truncated basis of single-particle orbitals, 
while we want to maintain the flexibility to describe the nuclei in any basis, or even through an ensemble of trajectories.  For brevity, we also suppress the subscript denoting the conditional dependence of $|\Phi\rangle$ on the generalized coordinates $Q$. In separating the center-of-mass coordinate, $\M^{\mu\nu}$ emerges in a $Q$-independent potentially off-diagonal form \cite{sutcliffe2000,requist2016b}.
Equations~(\ref{eq:EF:cm:chi}) and (\ref{eq:EF:cm:Phi}) are invariant under the linear coordinate transformation
\begin{align}
\tilde{Q}^{\mu} = T^{\mu}_{\nu} Q^{\nu} {,}
\end{align}
where $T^{\mu}_{\nu}$ is a $Q$-independent $O(3N-3)$ matrix.  However, if one transforms to a body-fixed frame, which requires a nonlinear coordinate transformation, the inverse mass tensor 
is no longer coordinate independent, and this motivates us to look at how the EF equations transform under general coordinate transformations.

The form of the EF equations in Eqs.~(\ref{eq:EF:cm:chi}) and (\ref{eq:EF:cm:Phi}), like those in Eqs.~(\ref{eq:EF:lab:chi}) and (\ref{eq:EF:lab:Phi}), is not invariant under the general coordinate transformation
\begin{align}
x^{\mu} = x^{\mu}(Q) {,}
\label{eq:coor:general}
\end{align}
where each new coordinate $x^{\mu}$ is given as a function of the $3N-3$ coordinates $Q^{\mu}$.  This is because when applying the coordinate transformation to the second derivative
\begin{align}
\f{\p}{\p Q^{\mu}} \f{\p}{\p Q^{\nu}} = \f{\p x^{\alpha}}{\p Q^{\mu}} \f{\p}{\p x^{\alpha}} \bigg( \f{\p x^{\beta}}{\p Q^{\nu}} \f{\p}{\p x^{\beta}} \bigg) {,}
\end{align}
an additional term arises from the action of $\p/\p x^{\alpha}$ on the $\p x^{\beta}/\p Q^{\nu}$ factor.

We now derive the EF equations in a coordinate-invariant form, starting from the nuclear kinetic energy of a state $|\Psi(Q,t)\rangle = \chi(Q,t) |\Phi(Q,t)\rangle$, 
\begin{align}
T_n &= \f{\hbar^2}{2} \int dQ \sqrt{\M} J_0 \p_{\mu} \big[\langle \Phi(Q,t)| \chi^*(Q,t) \big] \M^{\mu\nu} \p_{\nu} \big[\chi(Q,t) |\Phi(Q,t) \rangle \big] 
{,} 
\label{eq:Tn:functional}
\end{align}
where $\M=\det[\M_{\mu\nu}(Q)]$ and $J=\sqrt{\M} J_0$ is the Jacobian 
of the transformation $(r,R)\rightarrow (q,Q,\R_{cm})$; $J_0$ is the constant $M_{tot}^{3/2} (1-M_{elec}/M_{nucl})^{-3/2} \prod_{n=1}^N M_n^{-3/2}$, where $M_{elec}$ is the total electron mass, $M_{nucl}$ is the total nuclear mass, and $M_{tot}=M_{elec}+M_{nucl}$.
The inverse mass tensor $\M^{\mu\nu}=\M^{\mu\nu}(Q)$, now generalized to be $Q$-dependent, provides a metric on nuclear configuration space with dimensions of [mass]$^{-1}$.
The volume form on nuclear configuration space is $J dQ^1\wedge \ldots \wedge dQ^{3N-3}$.
The kinetic energy functional in Eq.~(\ref{eq:Tn:functional}) corresponds to the nuclear kinetic energy operator 
\begin{align}
\hat{T}_n = \f{1}{2} \f{1}{\sqrt{\M}} P_{\mu} \M^{\mu\nu} \sqrt{\M} P_{\nu} 
\label{eq:Tn}
\end{align}
in the Schr\"odinger equation for $|\Psi\rangle$.  $\hat{T}_n$ contains the Laplace-Beltrami operator used by Podolsky to express the kinetic energy operator in curvilinear coordinates \cite{podolsky1928}.
We now expand Eq.~(\ref{eq:Tn:functional}) as
\begin{align}
T_n 
&= \f{1}{2} \int dQ \sqrt{\M} J_0 \Big[ \M^{\mu\nu} \langle P_{\mu} \Phi | P_{\nu} \Phi \rangle |\chi|^2+ \langle \Phi | P_{\mu} \Phi \rangle \M^{\mu\nu} \chi^* P_{\nu} \chi - P_{\mu} \chi^* \chi \M^{\mu\nu} \langle \Phi | P_{\nu} \Phi \rangle - P_{\mu} \chi^* \M^{\mu\nu} P_{\nu} \chi \Big] 
\end{align}
and define the action 
\begin{align}
S[\chi,\Phi] = \int dt \int dQ \sqrt{\M} J_0 \chi^*(Q,t) \langle \Phi(Q,t) |i\hbar \p_t \big[ \chi(Q,t) \Phi(Q,t) \rangle \big] - \int dt (T_n + E^{BO}) {,}
\end{align}
where 
\begin{align}
E^{BO} = \int dQ \sqrt{\M} J_0 |\chi(Q)|^2 \langle \Phi | \hat{H}^{BO} | \Phi \rangle {.}
\end{align}
The Euler-Lagrange equation 
\begin{align}
\sqrt{\M} J_0 (i\hbar \p_t - A_0) \chi = \f{\delta T_n}{\delta \chi^*} - \p_{\mu} \f{\delta T_n}{\delta \p_{\mu}\chi^*} + \f{\delta E^{BO}}{\delta \chi^*} 
\label{eq:EF:chi:intermediate}
\end{align}
leads to the Schr\"odinger equation
\begin{align}
i\hbar D_t \chi = \f{1}{2} \f{1}{\sqrt{\M}} (P_{\mu}+A_{\mu}) \Big[ \sqrt{\M} \M^{\mu\nu} (P_{\nu}+A_{\nu}) \chi \Big] + (\mathcal{E}^{BO}+\mathcal{E}_{geo}) \chi {.}
\label{eq:EF:chi}
\end{align}
Since the operator in the first term on the right hand side is invariant to general gauge and coordinate transformations and the potentials $\mathcal{E}^{BO}$ and $\mathcal{E}_{geo}$ are also invariant, Eq.~(\ref{eq:EF:chi}) is the desired gauge- and coordinate-invariant equation for the marginal amplitude.
The kinetic energy operator can be expressed in terms of 
a gauge- and coordinate-covariant derivative $\nabla_{\mu}$ as (see App.~A) 
\begin{align}
-\f{\hbar^2}{2} \M^{\mu\nu} \nabla_{\mu} \nabla_{\nu} \chi = -\f{\hbar^2}{2} \M^{\mu\nu} D_{\mu}D_{\nu} \chi + \f{\hbar^2}{2} \M^{\mu\nu} \Pi^{\lambda}_{\mu\nu} D_{\lambda} \chi {,} 
\label{eq:Tn:christoffel}
\end{align}
where
\begin{align}
\Pi^{\lambda}_{\mu\nu} = \f{1}{2} \M^{\lambda\kappa} \bigg( \f{\p \M_{\mu\kappa}}{\p Q^{\nu}} + \f{\p \M_{\kappa\nu}}{\p Q^{\mu}} - \f{\p \M_{\mu\nu}}{\p Q^{\kappa}} \bigg) 
\label{eq:Pi}
\end{align}
is directly analogous to the Christoffel symbol $\Gamma^{\lambda}_{\mu\nu}$ in Riemannian geometry. 
We demonstrate the coordinate invariance of the kinetic energy operator in Eq.~(\ref{eq:Tn:christoffel}) in App.~B.
Finally, the manifestly gauge- and coordinate-invariant EF equation for the marginal amplitude is
\begin{align}
i\hbar D_t \chi = -\f{1}{2} \M^{\mu\nu} \nabla_{\mu} \nabla_{\nu} \chi + (\mathcal{E}^{BO}+\mathcal{E}_{geo}) \chi {.}
\label{eq:EF:chi:covariant}
\end{align}

The electronic Schr\"odinger-like equation obtained from the Euler-Lagrange equation
\begin{align}
\sqrt{\M} J_0 |\chi|^2 i\hbar\p_t |\Phi\rangle = \f{\delta T_n}{\delta \langle \Phi|} - \p_{\mu} \f{\delta T_n}{\delta \langle \p_{\mu}\Phi|} + \f{\delta E^{BO}}{\delta \langle \Phi|} 
\end{align}
is
\begin{align}
i\hbar D_t |\Phi\rangle &= \f{1}{2} \f{1}{\sqrt{\M}} (P_{\mu}-A_{\mu}) \Big[ \sqrt{\M} \M^{\mu\nu} (P_{\nu}-A_{\nu}) |\Phi\rangle \Big] + \f{\hbar^2}{2} \M^{\mu\nu} \Pi_{\mu\nu}^{\lambda} (\p_{\lambda}-iA_{\lambda}) |\Phi\rangle + \M^{\mu\nu} \f{(P_{\mu}+A_{\mu})\chi}{\chi} (P_{\nu}-A_{\nu}) |\Phi\rangle \nn \\
&\quad+ (\hat{H}^{BO} - \mathcal{E}^{BO} - \mathcal{E}_{geo}) |\Phi\rangle {.}
\label{eq:EF:Phi}
\end{align}
Using the gauge- and coordinate-covariant derivative $\nabla_{\mu}$, 
this can be written as
\begin{align}
i\hbar D_t |\Phi\rangle &= -\f{\hbar^2}{2} \M^{\mu\nu} \nabla_{\mu} \nabla_{\nu} |\Phi\rangle - \hbar^2 \M^{\mu\nu} \f{\nabla_{\mu}\chi}{\chi} \nabla_{\nu} |\Phi\rangle + (\hat{H}^{BO} - \mathcal{E}^{BO} - \mathcal{E}_{geo}) |\Phi\rangle {.}
\label{eq:EF:Phi:covariant}
\end{align}

In the special case that $\mathcal{M}_{\mu\nu}$ is $Q$-independent, 
the symbol $\Pi^{\lambda}_{\mu\nu}$ vanishes, $\nabla_{\mu}$ reduces to $\mathbf{D}_{\mu}$, and the EF equations reduce to those in Eqs.~(\ref{eq:EF:cm:chi}) and (\ref{eq:EF:cm:Phi}).  An important case in which the mass tensor is coordinate dependent arises in the transformation to orientational and shape coordinates by the definition of a body frame.
In this case, there is no choice of shape coordinates in which the mass tensor is coordinate independent
because the configuration (shape) space is non-Euclidean. 
After the transformation to 3 orientational coordinates and $3N-6$ shape coordinates $q^{\mu}$, the Hamiltonian for $N$ nuclei in the BO approximation takes the form \cite{littlejohn1997}
\begin{align}
H = \f{1}{2} \hat{\mathbf{L}} \cdot \mathcal{I}^{-1} \cdot \hat{\mathbf{L}} + \f{1}{2} (\hat{p}_{\mu} - \hat{\mathbf{L}} \cdot \mathbf{A}_{\mu}) \mathcal{M}^{\mu\nu} (\hat{p}_{\nu}-\hat{\mathbf{L}}\cdot \mathbf{A}_{\nu}) + \mathcal{E}_2 + \mathcal{E}_{BO} {,}
\label{eq:littlejohn}
\end{align}
where $\hat{\mathbf{L}}$ is the angular momentum operator in the body-fixed frame, $\hat{p}_{\mu} = -i\hbar\p/\p q^{\mu}$, 
\begin{align}
\mathcal{I}_{\alpha\beta}=\sum_{n=1}^{N-1}\big(|\vec{\rho}_n|^2 \delta_{\alpha\beta} - \rho_{n\alpha} \rho_{n\beta}\big), \quad \alpha,\beta=x,y,z 
\end{align}
is the moment of inertia tensor in terms of body-frame mass-weighted Jacobi coordinates $\vec{\rho}_n$, 
\begin{align}
\mathbf{A}_{\mu} = \mathcal{I}^{-1} \sum_{n=1}^{N-1} \vec{\rho}_n \times \f{\p\vec{\rho}_n}{\p q^{\mu}} {,}
\end{align}
and $\mathcal{E}_2=\mathcal{E}_2(q)$ is a potential that simplifies to \cite{watson1968} 
\begin{align}
-\f{\hbar^2}{8} \mathrm{Tr}\big[\mathcal{I}^{-1} + \mathbf{A}_{\mu} \mathcal{M}^{\mu\nu} \mathbf{A}_{\nu}^T\big] \label{eq:watson}
\end{align}
in the Eckart frame \cite{eckart1935}.
The Hamiltonian in Eq.~(\ref{eq:littlejohn}) is invariant to the choice of body frame, which is a type of gauge choice, due to the presence of the gauge potential $\mathbf{A}_{\mu}$, which is not to be confused with the vector potential in Eqs.~(\ref{eq:EF:lab:chi}) and (\ref{eq:EF:lab:Phi}).  In Eqs.~(\ref{eq:littlejohn}) and (\ref{eq:watson}), we have continued to use the notation $\mathcal{M}^{\mu\nu}$ for the inverse mass tensor.  It is now a dimensionless $q$-dependent metric tensor on shape space because the masses have been absorbed into the mass-weighted Jacobi coordinates.
A structure similar to that in Eq.~(\ref{eq:littlejohn}) arises when one performs a transformation to orientational and shape coordinates starting from the full Hamiltonian of electrons and nuclei without invoking the BO approximation.  We leave the derivation of the EF equations in this case for future work.

To highlight another aspect of the Riemannian geometry underpinning the EF equations in Eqs.~(\ref{eq:EF:chi:covariant}) and (\ref{eq:EF:Phi:covariant}), we write the geometric potential $\mathcal{E}_{geo}$ in terms of the Riemannian metric 
\begin{align}
g_{\mu\nu} &= \mathrm{Re} \langle (P_{\mu}-A_{\mu}) \Phi | (P_{\nu}-A_{\nu}) \Phi \rangle 
\label{eq:def:g}
\end{align}
as
\begin{align}
\mathcal{E}_{geo} = \f{1}{2} \M^{\mu\nu} g_{\mu\nu} {.}
\end{align}
As the contraction of contravariant and covariant tensors, $\mathcal{E}_{geo}$ is clearly invariant under general coordinate transformations.  The metric $g$ is called the quantum metric.  It is the pullback of the Fubini-Study metric on projective Hilbert space \cite{provost1980} to nuclear configuration space.  
Equation (\ref{eq:def:g}) defines a nonadiabatic quantum metric \cite{requist2016a,requist2016b,requist2017}, 
extending the well-known adiabatic metric defined in the BO approximation \cite{berry1989,berry1990,berry1993}.  
The quantum metric $g$ plays a crucial role in EF-based density functional theory \cite{requist2016b,li2018,requist2019b}, it determines the local-in-time error in the adiabatic approximation \cite{martinazzo2022}, and its equation of motion has been derived \cite{requist2022a}.
An analogous quantum metric appears in the factorization of an $n$-electron wavefunction into a one-electron marginal factor and an $(n-1)$-electron conditional factor \cite{schild2017,kocak2021}.  The quantum metric for cell-periodic Bloch functions defined on the Brillouin zone has recently found numerous applications in condensed matter physics.

\section{Quantum Christoffel symbol}

The conditional electronic state $|\Phi(Q)\rangle$, depending parametrically on the generalized nuclear coordinates $Q^{\mu}$, defines a mapping from configuration space $\mathcal{Q}$ to 
the space $\mathcal{P}$ of electronic states that obey the normalization $\langle \Phi(Q)|\Phi(Q)\rangle=1$ for all $Q$.  Let $\mathcal{Q}$ be $d$-dimensional and let $\mathcal{P}$ have real dimension $2n-1$, corresponding to an $n$-level electronic system. 

Consider the $D_{\mu}$-derivative of $|D_{\nu} \Phi \rangle$ and write it as
\begin{align}
D_{\mu} |D_{\nu} \Phi \rangle &= |\Phi\rangle \langle \Phi | D_{\mu} D_{\nu} \Phi \rangle + |D_{\lambda} \Phi\rangle \Upsilon^{\lambda}_{\mu\nu} + |e_a\rangle \Omega^a_{\mu\nu} {,}
\label{eq:DD}
\end{align}
where $\Upsilon^{\lambda}_{\mu\nu}$ is a rank-3 quantity, called the quantum Christoffel symbol of the second kind \cite{requist2022b}, $\Omega^a_{\mu\nu}$ is another rank-3 quantity, and $|e_a\rangle$ are basis vectors that span the subspace of the tangent space of $\mathcal{P}$ at $|\Phi\rangle$ that is orthogonal to all $|D_{\mu}\Phi\rangle$, i.e.~$\langle D_{\mu}\Phi|e_a\rangle= 0$ for all $\mu=1,\ldots,d$ and $a=1,\ldots,2n-2-d$.
Together, $|D_{\mu}\Phi\rangle$ and $|e_a\rangle$ form a complete 
basis for the tangent space of $\mathcal{P}$.  
Projecting Eq.~(\ref{eq:DD}) onto $|D_{\lambda}\Phi\rangle$ gives
\begin{align} 
\langle D_{\lambda} \Phi | D_{\mu} D_{\nu} \Phi \rangle &= \langle D_{\lambda} \Phi | D_{\kappa} \Phi\rangle \Upsilon^{\kappa}_{\mu\nu} {.}\label{eq:Upsilon:relation:precursor}
\end{align}
After defining the quantum geometric tensor \cite{provost1980,berry1989}
\begin{align}
h_{\lambda\kappa} &= \langle D_{\lambda} \Phi | D_{\kappa} \Phi \rangle
\end{align}
and the quantum Christoffel symbol of the first kind
\begin{align}
\Upsilon_{\lambda\mu\nu} &= \langle D_{\lambda} \Phi | D_{\mu} D_{\nu} \Phi \rangle {,}
\label{eq:Upsilon:definition}
\end{align}
Eq.~(\ref{eq:Upsilon:relation:precursor}) yields the relation
\begin{align}
\Upsilon_{\lambda\mu\nu} &= h_{\lambda\kappa} \Upsilon^{\kappa}_{\mu\nu} {.}
\label{eq:Upsilon:relation}
\end{align}
The real part of $\Upsilon_{\lambda\mu\nu}$ is equal to the Christoffel symbol of the first kind
\begin{align}
\Gamma_{\lambda\mu\nu} = \f{1}{2\hbar^2} \bigg( \f{\p g_{\lambda\mu}}{\p Q^{\nu}} + \f{\p g_{\lambda\nu}}{\p Q^{\mu}} - \f{\p g_{\mu\nu}}{\p Q^{\lambda}} \bigg) {,}
\end{align} 
while its imaginary part, $C_{\lambda\mu\nu}$, is a quantity whose physical significance has not been widely explored, although it appears in the equation of motion for $g_{\mu\nu}$ \cite{requist2022a}.

If $h_{\kappa\lambda}$ is invertible and its inverse is denoted $h^{\lambda\kappa}$, then we can write
\begin{align}
\Upsilon^{\lambda}_{\mu\nu} &= h^{\lambda\kappa} \Upsilon_{\kappa\mu\nu} {.}
\label{eq:Upsilon:12:from:Upsilon:03}
\end{align}
The symbol $\Upsilon^{\lambda}_{\mu\nu}$ defines a connection, which has been called the quantum covariant derivative $\hat{\nabla}$, in a Hermitian bundle over nuclear configuration space $\mathcal{Q}$ \cite{requist2022b}.  
The quantum covariant derivative $\hat{\nabla}$ has been used to derive a geometric version of adiabatic perturbation theory. 
It is important to clarify that the gauge- and coordinate-covariant derivative $\nabla$ defined in the preceding section is not the same as the quantum covariant derivative $\hat{\nabla}$.

The projection of the electronic EF equation in Eq.~(\ref{eq:EF:Phi}) onto $|\Phi\rangle$ gives a trivial identity.
The electronic EF equation lives in the tangent bundle of the space of normalized electronic functions.
%
%
Projecting Eq.~(\ref{eq:EF:Phi}) onto $|D_{\kappa} \Phi\rangle$ and $|e_b\rangle$, we obtain
\begin{align}
i \hbar \langle D_{\kappa} \Phi | D_t \Phi\rangle &= \langle D_{\kappa}\Phi | \hat{H}^{BO} | \Phi \rangle + \f{\hbar^2}{2} \M^{\mu\nu} (\Pi_{\mu\nu}^{\lambda}-\Upsilon^{\lambda}_{\mu\nu}) h_{\kappa\lambda} - \hbar^2 \M^{\mu\nu} \f{D_{\mu}\chi}{\chi} h_{\kappa\nu} \qquad \kappa=1,\ldots,d   
\label{eq:EF:elec:proj}
\end{align}
and
\begin{align}
i \hbar \langle e_b | D_t \Phi\rangle &= \langle e_b | \hat{H}^{BO} | \Phi \rangle - \f{\hbar^2}{2} \M^{\mu\nu} \Omega^{a}_{\mu\nu} \langle e_b | e_a\rangle  \qquad b=1,\ldots, 2n-2-d{.}
\label{eq:EF:elec:proj:2}
\end{align}
Equations~(\ref{eq:EF:elec:proj}) and (\ref{eq:EF:elec:proj:2}) are a set of $n-2$ equations that are equivalent to Eq.~(\ref{eq:EF:Phi}).  The quantum Christoffel symbol $\Upsilon^{\lambda}_{\mu\nu}$, in combination with $\Pi^{\lambda}_{\mu\nu}$, appears explicitly in Eq.~(\ref{eq:EF:elec:proj}), which might point the way to further understanding of its physical significance.

\section{Conclusions}

The exact factorization equations have been derived in a gauge- and coordinate-invariant form.  
Numerous different coordinate choices, such as Jacobi coordinates, hyperspherical coordinates and other curvilinear coordinates, have been used in molecular calculations. 
The most convenient choice depends on the problem under study.  For example, the best coordinates for describing vibrational spectroscopy are not the same as the best coordinates for molecular scattering calculations.
%
In some cases, using curvilinear coordinates is a matter of convenience, but in the transformation to rotational and vibrational coordinates, which is used e.g.~to treat angular momentum eigenstates, there is no choice, since the vibrational configuration space (shape space) is non-Euclidean.  
For this reason, it is desirable to derive the EF equations in a coordinate-invariant form. In doing so, we have made sure to preserve the gauge invariance of the equations.  Our analysis uncovers another role of Riemannian geometry in the EF method.

%
%





\begin{acknowledgments}
Funding from the European Research Council (ERC) under the European Union's Horizon 2020 research and innovation programme (grant agreement No. ERC-2017-AdG-788890) is gratefully acknowledged.
\end{acknowledgments}

\appendix



\section{Laplace-Beltrami operator}


In this appendix, we express the kinetic energy operator in Eq.~(\ref{eq:EF:chi}) in terms of the $\Pi^{\lambda}_{\mu\nu}$ symbol, and then in terms of the gauge- and coordinate-invariant derivative $\nabla_{\mu}$. 
We start from the first term on the right hand side of Eq.~(\ref{eq:EF:chi}):
\begin{align}
\f{1}{2} \f{1}{\sqrt{\M}} (P_{\mu}+A_{\mu}) \Big[ \sqrt{\M} \M^{\mu\nu} (P_{\nu}+A_{\nu}) \chi \Big] 
&= -\f{\hbar^2}{2} \M^{\mu\nu} (\p_{\mu}+iA_{\mu})(\p_{\nu}+iA_{\nu})\chi 
-\f{\hbar^2}{2} \p_{\mu} \M^{\mu\nu} (\p_{\nu}+iA_{\nu})\chi \nn \\
&\quad-\f{\hbar^2}{2} \f{\p_{\mu}\sqrt{\M}}{\sqrt{\M}} \M^{\mu\nu} (\p_{\nu}+iA_{\nu})\chi
{.} \label{eq:identity:1}
\end{align}
The second term on the right hand side of Eq.~(\ref{eq:identity:1}) is
\begin{align}
-\f{\hbar^2}{2} \p_{\mu} \M^{\mu\nu} (\p_{\nu}+iA_{\nu})\chi &= \f{\hbar^2}{2} \M^{\mu\lambda} \M^{\kappa\nu} \p_{\mu} \M_{\kappa\lambda} (\p_{\nu}+iA_{\nu}) \chi {.}
\end{align}
Changing dummy indices according to $\mu \rightarrow \beta$ and $\lambda \rightarrow \alpha$, we obtain
\begin{align}
\f{\hbar^2}{2} \M^{\beta\alpha} \M^{\kappa\nu} \p_{\beta} \M_{\kappa\alpha} (\p_{\nu}+iA_{\nu}) \chi
\end{align}
and recognize $\f{1}{2} \M^{\kappa\nu} \p_{\beta} \M_{\kappa\alpha}$ as the first term of 
\begin{align}
\Pi^{\nu}_{\alpha\beta} = \f{1}{2} \M^{\kappa\nu} \big( \p_{\beta} \M_{\alpha\kappa} + \p_{\alpha} \M_{\kappa\beta} - \p_{\kappa} \M_{\alpha\beta} \big) {.}
\end{align}
The first and second terms of $\Pi_{\alpha\beta}^{\nu}$ give equal contributions after contraction with $\M^{\alpha\beta}$.

After changing $\mu\rightarrow \kappa$, the third term on the right hand side of Eq.~(\ref{eq:identity:1}) can be written as
\begin{align}
-\f{\hbar^2}{2} \f{\p_{\kappa}\sqrt{\M}}{\sqrt{\M}} \M^{\kappa\nu} (\p_{\nu}+iA_{\nu})\chi &= -\f{\hbar^2}{4} \M^{\alpha\beta} \p_{\kappa} \M_{\alpha\beta} \M^{\kappa\nu} (\p_{\nu}+iA_{\nu})\chi {,}
\end{align}
where we recognize $-\f{1}{2} \M^{\kappa\nu} \p_{\kappa} \M_{\alpha\beta}$ as the last term of $\Pi_{\alpha\beta}^{\nu}$.
Hence, the sum of the second and third terms on the right hand side of Eq.~(\ref{eq:identity:1}) is equal to
\begin{align}
\f{\hbar^2}{2} \M^{\alpha\beta} \Pi_{\alpha\beta}^{\nu} (\p_{\nu}+iA_{\nu}) \chi {.}
\end{align}
This verifies that
\begin{align}
\f{1}{2} \f{1}{\sqrt{\M}} (P_{\mu}+A_{\mu}) \Big[ \sqrt{\M} \M^{\mu\nu} (P_{\nu}+A_{\nu}) \chi \Big] 
&= -\f{\hbar^2}{2} \M^{\mu\nu} (\p_{\mu}+iA_{\mu})(\p_{\nu}+iA_{\nu})\chi + \f{\hbar^2}{2} \M^{\mu\nu} \Pi_{\mu\nu}^{\lambda} (\p_{\lambda}+iA_{\lambda})\chi {.}
\end{align}
Equation~(\ref{eq:Tn:christoffel}) is verified after defining the gauge- and coordinate-covariant derivative by its action, i.e.~$\nabla_{\nu}\chi=D_{\nu}\chi$ and
\begin{align}
\nabla_{\mu} \nabla_{\nu} \chi = D_{\mu} D_{\nu} \chi - \Pi^{\lambda}_{\mu\nu} D_{\lambda}\chi {.}
\end{align}
The covariant derivative $\nabla_{\mu}$ acts the same way on $\nabla_{\nu} |\Phi\rangle$, except the definition of the derivative $D_{\mu}$ changes to $D_{\mu}=\p_{\mu}-iA_{\mu}$ due to the different gauge transformation property of $|\Phi\rangle$.


%


\section{Invariance of the kinetic energy}

Here we demonstrate the invariance of the kinetic energy under the transformation $Q^{\mu}\rightarrow \oQ^{\mu}=\oQ^{\mu}(Q)$.
\begin{align}
\hat{\o{T}}_n \o{\chi} &= -\f{\hbar^2}{2} \o{\M}^{\mu\nu} (\o{\p}_{\mu}+i\o{A}_{\mu})(\o{\p}_{\nu}+i\o{A}_{\nu})\o{\chi} + \f{\hbar^2}{2} \o{\M}^{\mu\nu} \o{\Pi}_{\mu\nu}^{\lambda} (\o{\p}_{\lambda}+i\o{A}_{\lambda})\o{\chi} \nn \\[0.2cm]
&= -\f{\hbar^2}{2} \f{\p \oQ^{\mu}}{\p Q^{\sigma}} \M^{\sigma\tau} \f{\p \oQ^{\nu}}{\p Q^{\tau}}  \f{\p Q^s}{\p \oQ^{\mu}} (\p_s+iA_s) \f{\p Q^{\rho}}{\p \oQ^{\nu}} (\p_{\rho}+iA_{\rho})\chi \nn \\
&\quad+\f{\hbar^2}{2} \f{\p \oQ^{\mu}}{\p Q^{\alpha}} \M^{\alpha\beta} \f{\p \oQ^{\nu}}{\p Q^{\beta}} \f{\p \oQ^{\lambda}}{\p Q^{\rho}} \Pi_{\sigma\tau}^{\rho} \f{\p Q^{\sigma}}{\p \oQ^{\mu}} \f{\p Q^{\tau}}{\p \oQ^{\nu}} \f{\p Q^r}{\p \oQ^{\lambda}} (\p_r+iA_r)\chi \nn \\
&\quad+\f{\hbar^2}{2} \f{\p \oQ^{\mu}}{\p Q^{\sigma}} \M^{\sigma\tau} \f{\p \oQ^{\nu}}{\p Q^{\tau}} \f{\p \oQ^{\lambda}}{\p Q^{\rho}} \f{\p^2 Q^{\rho}}{\p \oQ^{\mu} \p \oQ^{\nu}} \f{\p Q^r}{\p \oQ^{\lambda}} (\p_r+iA_r) \chi \nn \\[0.2cm]
&= -\f{\hbar^2}{2} \M^{\sigma\tau} (\p_{\sigma}+iA_{\sigma}) (\p_{\rho}+iA_{\rho}) \chi -\f{\hbar^2}{2} \M^{\sigma\tau} \f{\p \oQ^{\nu}}{\p Q^{\tau}} \bigg( \f{\p}{\p Q^{\sigma}} \f{\p Q^{\rho}}{\p \oQ^{\nu}} \bigg) (\p_{\rho}+iA_{\rho})\chi \nn \\
&\quad+\f{\hbar^2}{2} \M^{\sigma\tau} \Pi_{\sigma\tau}^{\rho} (\p_{\rho}+iA_{\rho})\chi \nn \\
&\quad+\f{\hbar^2}{2} \M^{\sigma\tau} \f{\p \oQ^{\mu}}{\p Q^{\sigma}} \f{\p \oQ^{\nu}}{\p Q^{\tau}} \f{\p^2 Q^{\rho}}{\p \oQ^{\mu} \p \oQ^{\nu}} (\p_{\rho}+iA_{\rho}) \chi \nn \\[0.2cm]
&= -\f{\hbar^2}{2} \M^{\sigma\tau} (\p_{\sigma}+iA_{\sigma}) (\p_{\rho}+iA_{\rho}) \chi +\f{\hbar^2}{2} \M^{\sigma\tau} \Pi_{\sigma\tau}^{\rho} (\p_{\rho}+iA_{\rho})\chi \nn \\
&= \hat{T}_n \chi {,}
\end{align}
where we used the transformation rule 
\begin{align}
\o{\Pi}^{\lambda}_{\mu\nu} = \f{\p \oQ^{\lambda}}{\p Q^{\rho}} \Pi_{\sigma\tau}^{\rho} \f{\p Q^{\sigma}}{\p \oQ^{\mu}} \f{\p Q^{\tau}}{\p \oQ^{\nu}} + \f{\p \oQ^{\lambda}}{\p Q^{\rho}} \f{\p^2 Q^{\rho}}{\p \oQ^{\mu} \p \oQ^{\nu}} {.}
\end{align}

\bibliography{coor-inv-2022b}

\begin{thebibliography}{49}%
\makeatletter
\providecommand \@ifxundefined [1]{%
 \@ifx{#1\undefined}
}%
\providecommand \@ifnum [1]{%
 \ifnum #1\expandafter \@firstoftwo
 \else \expandafter \@secondoftwo
 \fi
}%
\providecommand \@ifx [1]{%
 \ifx #1\expandafter \@firstoftwo
 \else \expandafter \@secondoftwo
 \fi
}%
\providecommand \natexlab [1]{#1}%
\providecommand \enquote  [1]{``#1''}%
\providecommand \bibnamefont  [1]{#1}%
\providecommand \bibfnamefont [1]{#1}%
\providecommand \citenamefont [1]{#1}%
\providecommand \href@noop [0]{\@secondoftwo}%
\providecommand \href [0]{\begingroup \@sanitize@url \@href}%
\providecommand \@href[1]{\@@startlink{#1}\@@href}%
\providecommand \@@href[1]{\endgroup#1\@@endlink}%
\providecommand \@sanitize@url [0]{\catcode `\\12\catcode `\$12\catcode
  `\&12\catcode `\#12\catcode `\^12\catcode `\_12\catcode `\%12\relax}%
\providecommand \@@startlink[1]{}%
\providecommand \@@endlink[0]{}%
\providecommand \url  [0]{\begingroup\@sanitize@url \@url }%
\providecommand \@url [1]{\endgroup\@href {#1}{\urlprefix }}%
\providecommand \urlprefix  [0]{URL }%
\providecommand \Eprint [0]{\href }%
\providecommand \doibase [0]{http://dx.doi.org/}%
\providecommand \selectlanguage [0]{\@gobble}%
\providecommand \bibinfo  [0]{\@secondoftwo}%
\providecommand \bibfield  [0]{\@secondoftwo}%
\providecommand \translation [1]{[#1]}%
\providecommand \BibitemOpen [0]{}%
\providecommand \bibitemStop [0]{}%
\providecommand \bibitemNoStop [0]{.\EOS\space}%
\providecommand \EOS [0]{\spacefactor3000\relax}%
\providecommand \BibitemShut  [1]{\csname bibitem#1\endcsname}%
\let\auto@bib@innerbib\@empty
\bibitem [{\citenamefont {Mead}\ and\ \citenamefont
  {Truhlar}(1979)}]{mead1979}%
  \BibitemOpen
  \bibfield  {author} {\bibinfo {author} {\bibfnamefont {C.~A.}\ \bibnamefont
  {Mead}}\ and\ \bibinfo {author} {\bibfnamefont {D.~G.}\ \bibnamefont
  {Truhlar}},\ }\href@noop {} {\bibfield  {journal} {\bibinfo  {journal} {J.
  Chem. Phys.}\ }\textbf {\bibinfo {volume} {70}},\ \bibinfo {pages} {2284}
  (\bibinfo {year} {1979})}\BibitemShut {NoStop}%
\bibitem [{\citenamefont {Gidopoulos}\ and\ \citenamefont
  {Gross}(2014)}]{gidopoulos2014}%
  \BibitemOpen
  \bibfield  {author} {\bibinfo {author} {\bibfnamefont {N.~I.}\ \bibnamefont
  {Gidopoulos}}\ and\ \bibinfo {author} {\bibfnamefont {E.~K.~U.}\ \bibnamefont
  {Gross}},\ }\href@noop {} {\bibfield  {journal} {\bibinfo  {journal} {Phil.
  Trans. Roy. Soc. A}\ }\textbf {\bibinfo {volume} {372}},\ \bibinfo {pages}
  {20130059} (\bibinfo {year} {2014})}\BibitemShut {NoStop}%
\bibitem [{\citenamefont {Hunter}(1975)}]{hunter1975}%
  \BibitemOpen
  \bibfield  {author} {\bibinfo {author} {\bibfnamefont {G.}~\bibnamefont
  {Hunter}},\ }\href@noop {} {\bibfield  {journal} {\bibinfo  {journal} {Int.
  J. Quantum Chem.}\ }\textbf {\bibinfo {volume} {9}},\ \bibinfo {pages} {237}
  (\bibinfo {year} {1975})}\BibitemShut {NoStop}%
\bibitem [{\citenamefont {Abedi}\ \emph {et~al.}(2010)\citenamefont {Abedi},
  \citenamefont {Maitra},\ and\ \citenamefont {Gross}}]{abedi2010}%
  \BibitemOpen
  \bibfield  {author} {\bibinfo {author} {\bibfnamefont {A.}~\bibnamefont
  {Abedi}}, \bibinfo {author} {\bibfnamefont {N.~T.}\ \bibnamefont {Maitra}}, \
  and\ \bibinfo {author} {\bibfnamefont {E.~K.~U.}\ \bibnamefont {Gross}},\
  }\href@noop {} {\bibfield  {journal} {\bibinfo  {journal} {Phys. Rev. Lett.}\
  }\textbf {\bibinfo {volume} {105}},\ \bibinfo {pages} {123002} (\bibinfo
  {year} {2010})}\BibitemShut {NoStop}%
\bibitem [{\citenamefont {Agostini}\ \emph {et~al.}(2013)\citenamefont
  {Agostini}, \citenamefont {Abedi}, \citenamefont {Suzuki},\ and\
  \citenamefont {Gross}}]{agostini2013}%
  \BibitemOpen
  \bibfield  {author} {\bibinfo {author} {\bibfnamefont {F.}~\bibnamefont
  {Agostini}}, \bibinfo {author} {\bibfnamefont {A.}~\bibnamefont {Abedi}},
  \bibinfo {author} {\bibfnamefont {Y.}~\bibnamefont {Suzuki}}, \ and\ \bibinfo
  {author} {\bibfnamefont {E.~K.~U.}\ \bibnamefont {Gross}},\ }\href@noop {}
  {\bibfield  {journal} {\bibinfo  {journal} {Mol. Phys.}\ }\textbf {\bibinfo
  {volume} {111}},\ \bibinfo {pages} {3625} (\bibinfo {year}
  {2013})}\BibitemShut {NoStop}%
\bibitem [{\citenamefont {Abedi}\ \emph {et~al.}(2014)\citenamefont {Abedi},
  \citenamefont {Agostini},\ and\ \citenamefont {Gross}}]{abedi2014}%
  \BibitemOpen
  \bibfield  {author} {\bibinfo {author} {\bibfnamefont {A.}~\bibnamefont
  {Abedi}}, \bibinfo {author} {\bibfnamefont {F.}~\bibnamefont {Agostini}}, \
  and\ \bibinfo {author} {\bibfnamefont {E.~K.~U.}\ \bibnamefont {Gross}},\
  }\href@noop {} {\bibfield  {journal} {\bibinfo  {journal} {Europhys. Lett.}\
  }\textbf {\bibinfo {volume} {106}},\ \bibinfo {pages} {33001} (\bibinfo
  {year} {2014})}\BibitemShut {NoStop}%
\bibitem [{\citenamefont {Min}\ \emph {et~al.}(2015)\citenamefont {Min},
  \citenamefont {Agostini},\ and\ \citenamefont {Gross}}]{min2015}%
  \BibitemOpen
  \bibfield  {author} {\bibinfo {author} {\bibfnamefont {S.~K.}\ \bibnamefont
  {Min}}, \bibinfo {author} {\bibfnamefont {F.}~\bibnamefont {Agostini}}, \
  and\ \bibinfo {author} {\bibfnamefont {E.~K.~U.}\ \bibnamefont {Gross}},\
  }\href@noop {} {\bibfield  {journal} {\bibinfo  {journal} {Phys. Rev. Lett.}\
  }\textbf {\bibinfo {volume} {115}},\ \bibinfo {pages} {073001} (\bibinfo
  {year} {2015})}\BibitemShut {NoStop}%
\bibitem [{\citenamefont {Agostini}\ \emph {et~al.}(2016)\citenamefont
  {Agostini}, \citenamefont {Min}, \citenamefont {Abedi},\ and\ \citenamefont
  {Gross}}]{agostini2016}%
  \BibitemOpen
  \bibfield  {author} {\bibinfo {author} {\bibfnamefont {F.}~\bibnamefont
  {Agostini}}, \bibinfo {author} {\bibfnamefont {S.~K.}\ \bibnamefont {Min}},
  \bibinfo {author} {\bibfnamefont {A.}~\bibnamefont {Abedi}}, \ and\ \bibinfo
  {author} {\bibfnamefont {E.~K.~U.}\ \bibnamefont {Gross}},\ }\href@noop {}
  {\bibfield  {journal} {\bibinfo  {journal} {J. Chem. Theory Comput.}\
  }\textbf {\bibinfo {volume} {12}},\ \bibinfo {pages} {2127} (\bibinfo {year}
  {2016})}\BibitemShut {NoStop}%
\bibitem [{\citenamefont {Agostini}\ \emph {et~al.}(2015)\citenamefont
  {Agostini}, \citenamefont {Abedi}, \citenamefont {Suzuki}, \citenamefont
  {Min}, \citenamefont {Maitra},\ and\ \citenamefont {Gross}}]{agostini2015}%
  \BibitemOpen
  \bibfield  {author} {\bibinfo {author} {\bibfnamefont {F.}~\bibnamefont
  {Agostini}}, \bibinfo {author} {\bibfnamefont {A.}~\bibnamefont {Abedi}},
  \bibinfo {author} {\bibfnamefont {Y.}~\bibnamefont {Suzuki}}, \bibinfo
  {author} {\bibfnamefont {S.~K.}\ \bibnamefont {Min}}, \bibinfo {author}
  {\bibfnamefont {N.~T.}\ \bibnamefont {Maitra}}, \ and\ \bibinfo {author}
  {\bibfnamefont {E.~K.~U.}\ \bibnamefont {Gross}},\ }\href@noop {} {\bibfield
  {journal} {\bibinfo  {journal} {J. Chem. Phys.}\ }\textbf {\bibinfo {volume}
  {142}},\ \bibinfo {pages} {084303} (\bibinfo {year} {2015})}\BibitemShut
  {NoStop}%
\bibitem [{\citenamefont {Min}\ \emph {et~al.}(2017)\citenamefont {Min},
  \citenamefont {Agostini}, \citenamefont {Tavernelli},\ and\ \citenamefont
  {Gross}}]{min2017}%
  \BibitemOpen
  \bibfield  {author} {\bibinfo {author} {\bibfnamefont {S.~K.}\ \bibnamefont
  {Min}}, \bibinfo {author} {\bibfnamefont {F.}~\bibnamefont {Agostini}},
  \bibinfo {author} {\bibfnamefont {I.}~\bibnamefont {Tavernelli}}, \ and\
  \bibinfo {author} {\bibfnamefont {E.~K.~U.}\ \bibnamefont {Gross}},\
  }\href@noop {} {\bibfield  {journal} {\bibinfo  {journal} {J. Phys. Chem.
  Lett.}\ }\textbf {\bibinfo {volume} {8}},\ \bibinfo {pages} {3048} (\bibinfo
  {year} {2017})}\BibitemShut {NoStop}%
\bibitem [{\citenamefont {Curchod}\ \emph {et~al.}(2018)\citenamefont
  {Curchod}, \citenamefont {Agostini},\ and\ \citenamefont
  {Tavernelli}}]{curchod2018}%
  \BibitemOpen
  \bibfield  {author} {\bibinfo {author} {\bibfnamefont {B.~F.~E.}\
  \bibnamefont {Curchod}}, \bibinfo {author} {\bibfnamefont {F.}~\bibnamefont
  {Agostini}}, \ and\ \bibinfo {author} {\bibfnamefont {I.}~\bibnamefont
  {Tavernelli}},\ }\href@noop {} {\bibfield  {journal} {\bibinfo  {journal}
  {Eur. Phys. J. B}\ }\textbf {\bibinfo {volume} {91}},\ \bibinfo {pages} {168}
  (\bibinfo {year} {2018})}\BibitemShut {NoStop}%
\bibitem [{\citenamefont {{Villaseco Arribas}}\ \emph
  {et~al.}(2022)\citenamefont {{Villaseco Arribas}}, \citenamefont {Agostini},\
  and\ \citenamefont {Maitra}}]{villaseco-arribas2022}%
  \BibitemOpen
  \bibfield  {author} {\bibinfo {author} {\bibfnamefont {E.}~\bibnamefont
  {{Villaseco Arribas}}}, \bibinfo {author} {\bibfnamefont {F.}~\bibnamefont
  {Agostini}}, \ and\ \bibinfo {author} {\bibfnamefont {N.~T.}\ \bibnamefont
  {Maitra}},\ }\href@noop {} {\bibfield  {journal} {\bibinfo  {journal}
  {Molecules}\ }\textbf {\bibinfo {volume} {27}},\ \bibinfo {pages} {4002}
  (\bibinfo {year} {2022})}\BibitemShut {NoStop}%
\bibitem [{\citenamefont {Requist}\ and\ \citenamefont
  {Gross}(2016)}]{requist2016b}%
  \BibitemOpen
  \bibfield  {author} {\bibinfo {author} {\bibfnamefont {R.}~\bibnamefont
  {Requist}}\ and\ \bibinfo {author} {\bibfnamefont {E.~K.~U.}\ \bibnamefont
  {Gross}},\ }\href@noop {} {\bibfield  {journal} {\bibinfo  {journal} {Phys.
  Rev. Lett.}\ }\textbf {\bibinfo {volume} {117}},\ \bibinfo {pages} {193001}
  (\bibinfo {year} {2016})}\BibitemShut {NoStop}%
\bibitem [{\citenamefont {Li}\ \emph {et~al.}(2018)\citenamefont {Li},
  \citenamefont {Requist},\ and\ \citenamefont {Gross}}]{li2018}%
  \BibitemOpen
  \bibfield  {author} {\bibinfo {author} {\bibfnamefont {C.}~\bibnamefont
  {Li}}, \bibinfo {author} {\bibfnamefont {R.}~\bibnamefont {Requist}}, \ and\
  \bibinfo {author} {\bibfnamefont {E.~K.~U.}\ \bibnamefont {Gross}},\
  }\href@noop {} {\bibfield  {journal} {\bibinfo  {journal} {J. Chem. Phys.}\
  }\textbf {\bibinfo {volume} {148}},\ \bibinfo {pages} {084110} (\bibinfo
  {year} {2018})}\BibitemShut {NoStop}%
\bibitem [{\citenamefont {Gonze}\ and\ \citenamefont {Lee}(1997)}]{gonze1997a}%
  \BibitemOpen
  \bibfield  {author} {\bibinfo {author} {\bibfnamefont {X.}~\bibnamefont
  {Gonze}}\ and\ \bibinfo {author} {\bibfnamefont {C.}~\bibnamefont {Lee}},\
  }\href@noop {} {\bibfield  {journal} {\bibinfo  {journal} {Phys. Rev. B}\
  }\textbf {\bibinfo {volume} {55}},\ \bibinfo {pages} {10355} (\bibinfo {year}
  {1997})}\BibitemShut {NoStop}%
\bibitem [{\citenamefont {Gonze}(1997)}]{gonze1997b}%
  \BibitemOpen
  \bibfield  {author} {\bibinfo {author} {\bibfnamefont {X.}~\bibnamefont
  {Gonze}},\ }\href@noop {} {\bibfield  {journal} {\bibinfo  {journal} {Phys.
  Rev. B}\ }\textbf {\bibinfo {volume} {55}},\ \bibinfo {pages} {10337}
  (\bibinfo {year} {1997})}\BibitemShut {NoStop}%
\bibitem [{\citenamefont {Baroni}\ \emph {et~al.}(2001)\citenamefont {Baroni},
  \citenamefont {{de Gironcoli}}, \citenamefont {{Dal Corso}},\ and\
  \citenamefont {Giannozzi}}]{baroni2001}%
  \BibitemOpen
  \bibfield  {author} {\bibinfo {author} {\bibfnamefont {S.}~\bibnamefont
  {Baroni}}, \bibinfo {author} {\bibfnamefont {S.}~\bibnamefont {{de
  Gironcoli}}}, \bibinfo {author} {\bibfnamefont {A.}~\bibnamefont {{Dal
  Corso}}}, \ and\ \bibinfo {author} {\bibfnamefont {P.}~\bibnamefont
  {Giannozzi}},\ }\href@noop {} {\bibfield  {journal} {\bibinfo  {journal}
  {Rev. Mod. Phys.}\ }\textbf {\bibinfo {volume} {73}},\ \bibinfo {pages} {515}
  (\bibinfo {year} {2001})}\BibitemShut {NoStop}%
\bibitem [{\citenamefont {Requist}\ \emph {et~al.}(2019)\citenamefont
  {Requist}, \citenamefont {Proetto},\ and\ \citenamefont
  {Gross}}]{requist2019b}%
  \BibitemOpen
  \bibfield  {author} {\bibinfo {author} {\bibfnamefont {R.}~\bibnamefont
  {Requist}}, \bibinfo {author} {\bibfnamefont {C.~R.}\ \bibnamefont
  {Proetto}}, \ and\ \bibinfo {author} {\bibfnamefont {E.~K.~U.}\ \bibnamefont
  {Gross}},\ }\href@noop {} {\bibfield  {journal} {\bibinfo  {journal} {Phys.
  Rev. B}\ }\textbf {\bibinfo {volume} {99}},\ \bibinfo {pages} {165136}
  (\bibinfo {year} {2019})}\BibitemShut {NoStop}%
\bibitem [{\citenamefont {Pellegrini}\ \emph {et~al.}(2022)\citenamefont
  {Pellegrini}, \citenamefont {Sanna}, \citenamefont {Requist},\ and\
  \citenamefont {Gross}}]{pellegrini2022}%
  \BibitemOpen
  \bibfield  {author} {\bibinfo {author} {\bibfnamefont {C.}~\bibnamefont
  {Pellegrini}}, \bibinfo {author} {\bibfnamefont {A.}~\bibnamefont {Sanna}},
  \bibinfo {author} {\bibfnamefont {R.}~\bibnamefont {Requist}}, \ and\
  \bibinfo {author} {\bibfnamefont {E.~K.~U.}\ \bibnamefont {Gross}},\
  }\href@noop {} {\bibfield  {journal} {\bibinfo  {journal} {J. Phys.: Condens.
  Matter}\ }\textbf {\bibinfo {volume} {34}},\ \bibinfo {pages} {183002}
  (\bibinfo {year} {2022})}\BibitemShut {NoStop}%
\bibitem [{\citenamefont {Rawlinson}\ and\ \citenamefont
  {Tronci}(2020)}]{rawlinson2020}%
  \BibitemOpen
  \bibfield  {author} {\bibinfo {author} {\bibfnamefont {J.~I.}\ \bibnamefont
  {Rawlinson}}\ and\ \bibinfo {author} {\bibfnamefont {C.}~\bibnamefont
  {Tronci}},\ }\href@noop {} {\bibfield  {journal} {\bibinfo  {journal} {Phys.
  Rev. A}\ }\textbf {\bibinfo {volume} {102}},\ \bibinfo {pages} {032811}
  (\bibinfo {year} {2020})}\BibitemShut {NoStop}%
\bibitem [{\citenamefont {Requist}\ \emph {et~al.}(2016)\citenamefont
  {Requist}, \citenamefont {Tandetzky},\ and\ \citenamefont
  {Gross}}]{requist2016a}%
  \BibitemOpen
  \bibfield  {author} {\bibinfo {author} {\bibfnamefont {R.}~\bibnamefont
  {Requist}}, \bibinfo {author} {\bibfnamefont {F.}~\bibnamefont {Tandetzky}},
  \ and\ \bibinfo {author} {\bibfnamefont {E.~K.~U.}\ \bibnamefont {Gross}},\
  }\href@noop {} {\bibfield  {journal} {\bibinfo  {journal} {Phys. Rev. A}\
  }\textbf {\bibinfo {volume} {93}},\ \bibinfo {pages} {042108} (\bibinfo
  {year} {2016})}\BibitemShut {NoStop}%
\bibitem [{\citenamefont {Hunter}(1986)}]{hunter1986}%
  \BibitemOpen
  \bibfield  {author} {\bibinfo {author} {\bibfnamefont {G.}~\bibnamefont
  {Hunter}},\ }\href@noop {} {\bibfield  {journal} {\bibinfo  {journal} {Int.
  J. Quant. Chem.}\ }\textbf {\bibinfo {volume} {29}},\ \bibinfo {pages} {197}
  (\bibinfo {year} {1986})}\BibitemShut {NoStop}%
\bibitem [{\citenamefont {Schild}\ and\ \citenamefont
  {Gross}(2017)}]{schild2017}%
  \BibitemOpen
  \bibfield  {author} {\bibinfo {author} {\bibfnamefont {A.}~\bibnamefont
  {Schild}}\ and\ \bibinfo {author} {\bibfnamefont {E.~K.~U.}\ \bibnamefont
  {Gross}},\ }\href@noop {} {\bibfield  {journal} {\bibinfo  {journal} {Phys.
  Rev. Lett.}\ }\textbf {\bibinfo {volume} {118}},\ \bibinfo {pages} {163202}
  (\bibinfo {year} {2017})}\BibitemShut {NoStop}%
\bibitem [{\citenamefont {Koc\'ak}\ and\ \citenamefont
  {Schild}(2020)}]{kocak2020}%
  \BibitemOpen
  \bibfield  {author} {\bibinfo {author} {\bibfnamefont {J.}~\bibnamefont
  {Koc\'ak}}\ and\ \bibinfo {author} {\bibfnamefont {A.}~\bibnamefont
  {Schild}},\ }\href@noop {} {\bibfield  {journal} {\bibinfo  {journal} {Phys.
  Rev. Research}\ }\textbf {\bibinfo {volume} {2}},\ \bibinfo {pages} {043365}
  (\bibinfo {year} {2020})}\BibitemShut {NoStop}%
\bibitem [{\citenamefont {Koc\'ak}\ \emph {et~al.}(2021)\citenamefont
  {Koc\'ak}, \citenamefont {Kraisler},\ and\ \citenamefont
  {Schild}}]{kocak2021}%
  \BibitemOpen
  \bibfield  {author} {\bibinfo {author} {\bibfnamefont {J.}~\bibnamefont
  {Koc\'ak}}, \bibinfo {author} {\bibfnamefont {E.}~\bibnamefont {Kraisler}}, \
  and\ \bibinfo {author} {\bibfnamefont {A.}~\bibnamefont {Schild}},\
  }\href@noop {} {\bibfield  {journal} {\bibinfo  {journal} {J. Phys. Chem.
  Lett.}\ }\textbf {\bibinfo {volume} {12}},\ \bibinfo {pages} {3204} (\bibinfo
  {year} {2021})}\BibitemShut {NoStop}%
\bibitem [{\citenamefont {Hoffmann}\ \emph {et~al.}(2018)\citenamefont
  {Hoffmann}, \citenamefont {Appel}, \citenamefont {Rubio},\ and\ \citenamefont
  {Maitra}}]{hoffmann2018}%
  \BibitemOpen
  \bibfield  {author} {\bibinfo {author} {\bibfnamefont {N.~M.}\ \bibnamefont
  {Hoffmann}}, \bibinfo {author} {\bibfnamefont {H.}~\bibnamefont {Appel}},
  \bibinfo {author} {\bibfnamefont {A.}~\bibnamefont {Rubio}}, \ and\ \bibinfo
  {author} {\bibfnamefont {N.~T.}\ \bibnamefont {Maitra}},\ }\href@noop {}
  {\bibfield  {journal} {\bibinfo  {journal} {Eur. Phys. J. B}\ }\textbf
  {\bibinfo {volume} {91}},\ \bibinfo {pages} {180} (\bibinfo {year}
  {2018})}\BibitemShut {NoStop}%
\bibitem [{\citenamefont {Gonze}\ \emph {et~al.}(2018)\citenamefont {Gonze},
  \citenamefont {Zhou},\ and\ \citenamefont {Reining}}]{gonze2018}%
  \BibitemOpen
  \bibfield  {author} {\bibinfo {author} {\bibfnamefont {X.}~\bibnamefont
  {Gonze}}, \bibinfo {author} {\bibfnamefont {J.~S.}\ \bibnamefont {Zhou}}, \
  and\ \bibinfo {author} {\bibfnamefont {L.}~\bibnamefont {Reining}},\
  }\href@noop {} {\bibfield  {journal} {\bibinfo  {journal} {Eur. Phys. J. B}\
  }\textbf {\bibinfo {volume} {91}},\ \bibinfo {pages} {224} (\bibinfo {year}
  {2018})}\BibitemShut {NoStop}%
\bibitem [{\citenamefont {Lacombe}\ and\ \citenamefont
  {Maitra}(2020)}]{lacombe2020}%
  \BibitemOpen
  \bibfield  {author} {\bibinfo {author} {\bibfnamefont {L.}~\bibnamefont
  {Lacombe}}\ and\ \bibinfo {author} {\bibfnamefont {N.~T.}\ \bibnamefont
  {Maitra}},\ }\href@noop {} {\bibfield  {journal} {\bibinfo  {journal} {Phys.
  Rev. Lett.}\ }\textbf {\bibinfo {volume} {124}},\ \bibinfo {pages} {206401}
  (\bibinfo {year} {2020})}\BibitemShut {NoStop}%
\bibitem [{\citenamefont {Requist}\ and\ \citenamefont
  {Gross}(2021)}]{requist2021}%
  \BibitemOpen
  \bibfield  {author} {\bibinfo {author} {\bibfnamefont {R.}~\bibnamefont
  {Requist}}\ and\ \bibinfo {author} {\bibfnamefont {E.~K.~U.}\ \bibnamefont
  {Gross}},\ }\href@noop {} {\bibfield  {journal} {\bibinfo  {journal} {Phys.
  Rev. Lett.}\ }\textbf {\bibinfo {volume} {127}},\ \bibinfo {pages} {116401}
  (\bibinfo {year} {2021})}\BibitemShut {NoStop}%
\bibitem [{\citenamefont {Aharonov}\ and\ \citenamefont
  {Bohm}(1959)}]{aharonov1959}%
  \BibitemOpen
  \bibfield  {author} {\bibinfo {author} {\bibfnamefont {Y.}~\bibnamefont
  {Aharonov}}\ and\ \bibinfo {author} {\bibfnamefont {D.}~\bibnamefont
  {Bohm}},\ }\href@noop {} {\bibfield  {journal} {\bibinfo  {journal} {Phys.
  Rev.}\ }\textbf {\bibinfo {volume} {115}},\ \bibinfo {pages} {485} (\bibinfo
  {year} {1959})}\BibitemShut {NoStop}%
\bibitem [{\citenamefont {Mead}(1980{\natexlab{a}})}]{mead1980a}%
  \BibitemOpen
  \bibfield  {author} {\bibinfo {author} {\bibfnamefont {C.~A.}\ \bibnamefont
  {Mead}},\ }\href@noop {} {\bibfield  {journal} {\bibinfo  {journal} {J. Chem.
  Phys.}\ }\textbf {\bibinfo {volume} {72}},\ \bibinfo {pages} {3839} (\bibinfo
  {year} {1980}{\natexlab{a}})}\BibitemShut {NoStop}%
\bibitem [{\citenamefont {Longuet-Higgins}\ \emph {et~al.}(1958)\citenamefont
  {Longuet-Higgins}, \citenamefont {\"Opik}, \citenamefont {Pryce},\ and\
  \citenamefont {Sack}}]{longuet-higgins1958}%
  \BibitemOpen
  \bibfield  {author} {\bibinfo {author} {\bibfnamefont {H.~C.}\ \bibnamefont
  {Longuet-Higgins}}, \bibinfo {author} {\bibfnamefont {U.}~\bibnamefont
  {\"Opik}}, \bibinfo {author} {\bibfnamefont {M.~H.~L.}\ \bibnamefont
  {Pryce}}, \ and\ \bibinfo {author} {\bibfnamefont {R.~A.}\ \bibnamefont
  {Sack}},\ }\href@noop {} {\bibfield  {journal} {\bibinfo  {journal} {Proc. R.
  Soc. London, Ser. A}\ }\textbf {\bibinfo {volume} {244}},\ \bibinfo {pages}
  {1} (\bibinfo {year} {1958})}\BibitemShut {NoStop}%
\bibitem [{\citenamefont {Mead}(1980{\natexlab{b}})}]{mead1980b}%
  \BibitemOpen
  \bibfield  {author} {\bibinfo {author} {\bibfnamefont {C.~A.}\ \bibnamefont
  {Mead}},\ }\href@noop {} {\bibfield  {journal} {\bibinfo  {journal} {Chem.
  Phys.}\ }\textbf {\bibinfo {volume} {43}},\ \bibinfo {pages} {33} (\bibinfo
  {year} {1980}{\natexlab{b}})}\BibitemShut {NoStop}%
\bibitem [{\citenamefont {Berry}(1984)}]{berry1984}%
  \BibitemOpen
  \bibfield  {author} {\bibinfo {author} {\bibfnamefont {M.~V.}\ \bibnamefont
  {Berry}},\ }\href@noop {} {\bibfield  {journal} {\bibinfo  {journal} {Proc.
  Roy. Soc. Lond. A}\ }\textbf {\bibinfo {volume} {392}},\ \bibinfo {pages}
  {45} (\bibinfo {year} {1984})}\BibitemShut {NoStop}%
\bibitem [{\citenamefont {Min}\ \emph {et~al.}(2014)\citenamefont {Min},
  \citenamefont {Abedi}, \citenamefont {Kim},\ and\ \citenamefont
  {Gross}}]{min2014}%
  \BibitemOpen
  \bibfield  {author} {\bibinfo {author} {\bibfnamefont {S.~K.}\ \bibnamefont
  {Min}}, \bibinfo {author} {\bibfnamefont {A.}~\bibnamefont {Abedi}}, \bibinfo
  {author} {\bibfnamefont {K.~S.}\ \bibnamefont {Kim}}, \ and\ \bibinfo
  {author} {\bibfnamefont {E.~K.~U.}\ \bibnamefont {Gross}},\ }\href@noop {}
  {\bibfield  {journal} {\bibinfo  {journal} {Phys. Rev. Lett.}\ }\textbf
  {\bibinfo {volume} {113}},\ \bibinfo {pages} {263004} (\bibinfo {year}
  {2014})}\BibitemShut {NoStop}%
\bibitem [{\citenamefont {Sutcliffe}(2000)}]{sutcliffe2000}%
  \BibitemOpen
  \bibfield  {author} {\bibinfo {author} {\bibfnamefont {B.~T.}\ \bibnamefont
  {Sutcliffe}},\ }\href@noop {} {\bibfield  {journal} {\bibinfo  {journal}
  {Adv. Chem. Phys.}\ }\textbf {\bibinfo {volume} {114}},\ \bibinfo {pages}
  {97} (\bibinfo {year} {2000})}\BibitemShut {NoStop}%
\bibitem [{\citenamefont {Podolsky}(1928)}]{podolsky1928}%
  \BibitemOpen
  \bibfield  {author} {\bibinfo {author} {\bibfnamefont {B.}~\bibnamefont
  {Podolsky}},\ }\href@noop {} {\bibfield  {journal} {\bibinfo  {journal}
  {Phys. Rev.}\ }\textbf {\bibinfo {volume} {32}},\ \bibinfo {pages} {812}
  (\bibinfo {year} {1928})}\BibitemShut {NoStop}%
\bibitem [{\citenamefont {Littlejohn}\ and\ \citenamefont
  {Reinsch}(1997)}]{littlejohn1997}%
  \BibitemOpen
  \bibfield  {author} {\bibinfo {author} {\bibfnamefont {R.~G.}\ \bibnamefont
  {Littlejohn}}\ and\ \bibinfo {author} {\bibfnamefont {M.}~\bibnamefont
  {Reinsch}},\ }\href@noop {} {\bibfield  {journal} {\bibinfo  {journal} {Rev.
  Mod. Phys.}\ }\textbf {\bibinfo {volume} {69}},\ \bibinfo {pages} {213}
  (\bibinfo {year} {1997})}\BibitemShut {NoStop}%
\bibitem [{\citenamefont {Watson}(1968)}]{watson1968}%
  \BibitemOpen
  \bibfield  {author} {\bibinfo {author} {\bibfnamefont {J.~K.~G.}\
  \bibnamefont {Watson}},\ }\href@noop {} {\bibfield  {journal} {\bibinfo
  {journal} {Molec. Phys.}\ }\textbf {\bibinfo {volume} {15}},\ \bibinfo
  {pages} {479} (\bibinfo {year} {1968})}\BibitemShut {NoStop}%
\bibitem [{\citenamefont {Eckart}(1935)}]{eckart1935}%
  \BibitemOpen
  \bibfield  {author} {\bibinfo {author} {\bibfnamefont {C.}~\bibnamefont
  {Eckart}},\ }\href@noop {} {\bibfield  {journal} {\bibinfo  {journal} {Phys.
  Rev.}\ }\textbf {\bibinfo {volume} {47}},\ \bibinfo {pages} {552} (\bibinfo
  {year} {1935})}\BibitemShut {NoStop}%
\bibitem [{\citenamefont {Provost}\ and\ \citenamefont
  {Vallee}(1980)}]{provost1980}%
  \BibitemOpen
  \bibfield  {author} {\bibinfo {author} {\bibfnamefont {J.~P.}\ \bibnamefont
  {Provost}}\ and\ \bibinfo {author} {\bibfnamefont {G.}~\bibnamefont
  {Vallee}},\ }\href@noop {} {\bibfield  {journal} {\bibinfo  {journal}
  {Commun. Math. Phys.}\ }\textbf {\bibinfo {volume} {76}},\ \bibinfo {pages}
  {289} (\bibinfo {year} {1980})}\BibitemShut {NoStop}%
\bibitem [{\citenamefont {R.~Requist}\ and\ \citenamefont
  {Gross}(2017)}]{requist2017}%
  \BibitemOpen
  \bibfield  {author} {\bibinfo {author} {\bibfnamefont {C.~R.~P.}\
  \bibnamefont {R.~Requist}}\ and\ \bibinfo {author} {\bibfnamefont {E.~K.~U.}\
  \bibnamefont {Gross}},\ }\href@noop {} {\bibfield  {journal} {\bibinfo
  {journal} {Phys. Rev. A}\ }\textbf {\bibinfo {volume} {96}},\ \bibinfo
  {pages} {062503} (\bibinfo {year} {2017})}\BibitemShut {NoStop}%
\bibitem [{\citenamefont {Berry}(1989)}]{berry1989}%
  \BibitemOpen
  \bibfield  {author} {\bibinfo {author} {\bibfnamefont {M.~V.}\ \bibnamefont
  {Berry}},\ }\enquote {\bibinfo {title} {The quantum phase, five years
  after},}\ \ (\bibinfo {year} {1989})\ pp.\ \bibinfo {pages} {7--28},\
  \bibinfo {note} {in ref.~\cite{shapere1989}}\BibitemShut {NoStop}%
\bibitem [{\citenamefont {Berry}\ and\ \citenamefont {Lim}(1990)}]{berry1990}%
  \BibitemOpen
  \bibfield  {author} {\bibinfo {author} {\bibfnamefont {M.~V.}\ \bibnamefont
  {Berry}}\ and\ \bibinfo {author} {\bibfnamefont {R.}~\bibnamefont {Lim}},\
  }\href@noop {} {\bibfield  {journal} {\bibinfo  {journal} {J. Phys. A: Math.
  Gen}\ }\textbf {\bibinfo {volume} {23}},\ \bibinfo {pages} {L655} (\bibinfo
  {year} {1990})}\BibitemShut {NoStop}%
\bibitem [{\citenamefont {Berry}\ and\ \citenamefont
  {Robbins}(1993)}]{berry1993}%
  \BibitemOpen
  \bibfield  {author} {\bibinfo {author} {\bibfnamefont {M.~V.}\ \bibnamefont
  {Berry}}\ and\ \bibinfo {author} {\bibfnamefont {J.~M.}\ \bibnamefont
  {Robbins}},\ }\href@noop {} {\bibfield  {journal} {\bibinfo  {journal} {Proc.
  R. Soc. Lond. A}\ }\textbf {\bibinfo {volume} {442}},\ \bibinfo {pages} {641}
  (\bibinfo {year} {1993})}\BibitemShut {NoStop}%
\bibitem [{\citenamefont {Martinazzo}\ and\ \citenamefont
  {Burghardt}(2022)}]{martinazzo2022}%
  \BibitemOpen
  \bibfield  {author} {\bibinfo {author} {\bibfnamefont {R.}~\bibnamefont
  {Martinazzo}}\ and\ \bibinfo {author} {\bibfnamefont {I.}~\bibnamefont
  {Burghardt}},\ }\href@noop {} {\bibfield  {journal} {\bibinfo  {journal}
  {Phys. Rev. A}\ }\textbf {\bibinfo {volume} {105}},\ \bibinfo {pages}
  {052215} (\bibinfo {year} {2022})}\BibitemShut {NoStop}%
\bibitem [{\citenamefont {Requist}\ \emph {et~al.}(2022)\citenamefont
  {Requist}, \citenamefont {Li},\ and\ \citenamefont {Gross}}]{requist2022a}%
  \BibitemOpen
  \bibfield  {author} {\bibinfo {author} {\bibfnamefont {R.}~\bibnamefont
  {Requist}}, \bibinfo {author} {\bibfnamefont {C.}~\bibnamefont {Li}}, \ and\
  \bibinfo {author} {\bibfnamefont {E.~K.~U.}\ \bibnamefont {Gross}},\
  }\href@noop {} {\bibfield  {journal} {\bibinfo  {journal}
  {Phil.~Trans.~R.~Soc.~A}\ }\textbf {\bibinfo {volume} {380}},\ \bibinfo
  {pages} {20200383} (\bibinfo {year} {2022})}\BibitemShut {NoStop}%
\bibitem [{\citenamefont {Requist}(2022)}]{requist2022b}%
  \BibitemOpen
  \bibfield  {author} {\bibinfo {author} {\bibfnamefont {R.}~\bibnamefont
  {Requist}},\ }\href@noop {} {}\bibinfo {howpublished} {arxiv:2206.01716}
  (\bibinfo {year} {2022})\BibitemShut {NoStop}%
\bibitem [{\citenamefont {Shapere}\ and\ \citenamefont
  {Wilczek}(1989)}]{shapere1989}%
  \BibitemOpen
  \bibinfo {editor} {\bibfnamefont {A.}~\bibnamefont {Shapere}}\ and\ \bibinfo
  {editor} {\bibfnamefont {F.}~\bibnamefont {Wilczek}},\ eds.,\ \href@noop {}
  {\emph {\bibinfo {title} {Geometric phases in physics}}}\ (\bibinfo
  {publisher} {World Scientific, Singapore},\ \bibinfo {year}
  {1989})\BibitemShut {NoStop}%
\end{thebibliography}%

\end{document}